\documentclass[lettersize,journal]{IEEEtran}
\IEEEoverridecommandlockouts

\usepackage[table]{xcolor}
\usepackage{cite}
\usepackage{amsmath,amssymb,amsfonts}
\newcommand{\ignore}[1]{ }
\usepackage{algorithmic}    
\usepackage{graphicx}
\usepackage{textcomp}
\usepackage{xcolor}
\usepackage{float}
\usepackage{multirow}
\usepackage{lipsum}
\usepackage{listings}
\usepackage{color}
\usepackage{amsmath}
\definecolor{dkgreen}{rgb}{0,0.6,0}
\definecolor{gray}{rgb}{0.5,0.5,0.5}
\definecolor{mauve}{rgb}{0.58,0,0.82}
\usepackage{enumitem}

\usepackage{adjustbox}
\usepackage{array}

\usepackage{pifont}
\newcommand{\cmark}{\ding{51}}%
\newcommand{\xmark}{\ding{55}}%

\usepackage{pifont}
\newcolumntype{P}[1]{>{\centering\arraybackslash}p{#1}}

\newcounter{mysfig}

\renewcommand\themysfig{\theFig.(\alph{mysfig})}
\makeatletter
\newcommand\Scaption[1]{%
\vskip.5\abovecaptionskip
  \sbox\@tempboxa{\small~#1}%
  \ifdim \wd\@tempboxa >\hsize
    \small\themysfig~#1\par
  \else
    \global \@minipagefalse
    \hb@xt@\hsize{\hfil\box\@tempboxa\hfil}%
  \fi
  \vskip\belowcaptionskip}
\makeatother
\usepackage{array}

\usepackage{balance}
\usepackage{hyperref}

\hypersetup{
  colorlinks,
  allcolors=blue
  citebordercolor=blue,
  citecolor=blue,
  linkcolor=blue,
  urlcolor=blue}

\usepackage{courier}

\usepackage{ragged2e}

\usepackage[prependcaption, textsize=footnotesize]{todonotes}
\usepackage{todonotes} 
\newcommand{\AmirHosein}[1]{\todo[inline,linecolor=black,backgroundcolor=orange!30,bordercolor=black]{AmirHosein: #1}}

\usepackage{array} 

\usepackage{tikz,xcolor,hyperref}

\definecolor{lime}{HTML}{A6CE39}
\DeclareRobustCommand{\orcidicon}{%
	\begin{tikzpicture}
	\draw[lime, fill=lime] (0,0) 
	circle [radius=0.16] 
	node[white] {{\fontfamily{qag}\selectfont \tiny ID}};
	\draw[white, fill=white] (-0.0625,0.095) 
	circle [radius=0.007];
	\end{tikzpicture}
	\hspace{-2mm}
}

\foreach \x in {A, ..., Z}{%
	\expandafter\xdef\csname orcid\x\endcsname{\noexpand\href{https://orcid.org/\csname orcidauthor\x\endcsname}{\noexpand\orcidicon}}
}

\begin{document}
\title{\huge Machine Learning-Driven Open-Source Framework for Assessing QoE in Multimedia Networks}

\author{Parsa H. S. Panahi\orcidB{}, Amir H. Jalilvand \orcidA{},   Abolfazl Diyanat \orcidC{}, \textit{Member, IEEE}

\vspace{-1.5em}

\thanks{The implementation of this research is available as open source on GitHub \cite{f4-8}, enabling other researchers and developers to access the project source code, make improvements, or adapt it to their specific needs.}
\thanks{The authors are affiliated with the School of Computer Engineering at Iran University of Science and Technology,  Tehran, Iran. Email: \{Parsa\_shariat, jalilvand\_a, a.diyanat\}@iust.ac.ir}}

\maketitle

\begin{abstract}

The Internet is integral to modern life, influencing communication, business, and lifestyles globally. As dependence on Internet services grows, the demand for high-quality service delivery increases. Service providers must maintain high standards of quality of service  and quality of experience (QoE) to ensure user satisfaction. QoE, which reflects user satisfaction with service quality, is a key metric for multimedia services, yet it is challenging to measure due to its subjective nature and the complexities of real-time feedback.
This paper introduces a machine learning-based framework for objectively assessing QoE in multimedia networks. The open-source framework complies with the ITU-T P.1203 standard. It automates data collection and user satisfaction prediction using key network parameters such as delay, jitter, packet loss, bitrate, and throughput. Using a dataset of over 20,000 records from various network conditions, the Random Forest model predicts the mean opinion score with 95.8\% accuracy.

Our framework addresses the limitations of existing QoE models by integrating real-time data collection, machine learning predictions, and adherence to international standards. This approach enhances QoE evaluation accuracy and allows dynamic network resource management, optimizing performance and cost-efficiency. Its open-source nature encourages adaptation and extension for various multimedia services.
The findings significantly affect the telecommunications industry in managing and optimizing multimedia services. The network-centric QoE prediction of the framework offers a scalable solution to improve user satisfaction without the need for content-specific data. Future enhancements could include advanced machine learning models and broader applicability to digital services. This research contributes a practical, standardized tool for QoE assessment across diverse networks and platforms.

\end{abstract}


\section{Introduction}


\ignore{
\begin{Fig.}
    \centering
    \includegraphics[width=0.9\linewidth]{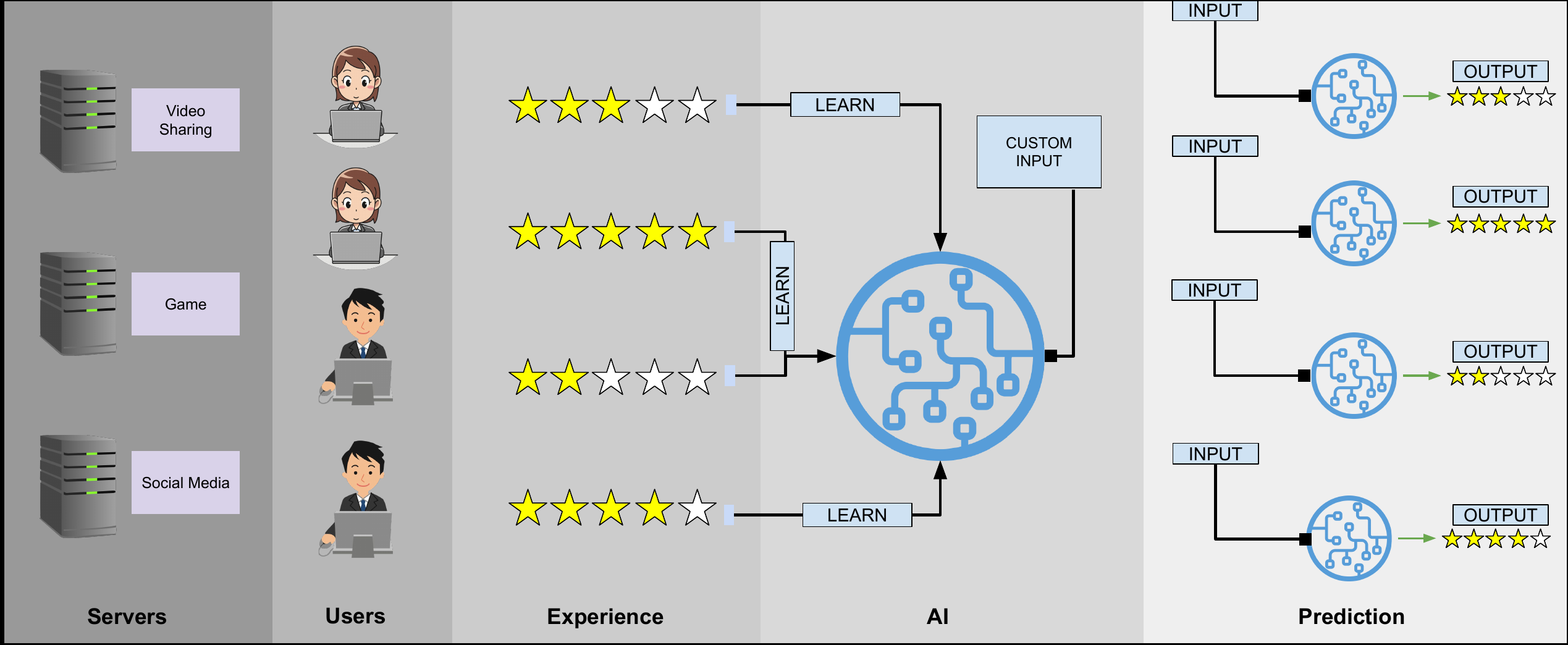}

    \AmirHosein{We don't have a cross reference to this image + I suggest yo remove background of the image in this work}
    \caption{The image demonstrates an end-to-end system where servers deliver services such as video sharing, gaming, and social media to users. Users rate their experiences, and these ratings, along with other inputs like network conditions, are fed into an AI model. The AI model learns from this data to predict Quality of Experience (QoE) scores independently \cite{mine}.}
    \label{fig:all-in-one}
\end{Fig.}
}

\IEEEPARstart{E}{xcellent} user experiences are fundamental in modern communication systems. This requires ensuring seamless interactions, intuitive designs, and dependable systems. Such efforts contribute to increased customer satisfaction and loyalty, enabling businesses to succeed in competitive markets. Understanding and improving user experiences is vital for maintaining a competitive advantage and meeting customer expectations \cite{mine, arellano2024survey}. Over the years, extensive research has been conducted on quality of experience (QoE), resulting in the development of various QoE assessment frameworks \cite{sr-8, sr-11, sr-1, sr-12, sr-20}. For instance, \texttt{YoMoApp} \cite{sr-1, sr-12} evaluates key performance indicators (KPIs) for YouTube video QoE in mobile networks. As described in \cite{sr-8}, a framework using a video server, network simulator, and receiver examines the impact of network parameters on IPTV user QoE. Reference \cite{sr-11} introduces a tool that enhances mobile network performance by integrating subjective user QoE data with objective network parameters. In \cite{sr-20}, a proactive LTE network management framework based on user QoE is proposed, which utilizes ML (ML) algorithms to predict QoE parameters using extensive cell-level network performance data.

\begin{table*}
\centering
\caption{Comparison of various studies on QoE assessment and management in multimedia services. The table outlines the main features addressed by each study, such as data collection tools, MOS calculation methods, consideration of human factors, ML techniques,  compliance with the ITU P.1203 standard, and helpfulness for resource allocation (RA). Obj: objective, Subj.: subjective }
\begin{tabular}{|P{1.6cm}|P{0.5cm}|p{4.5cm}|P{0.7cm}|P{1cm}|P{0.6cm}|P{0.6cm}|P{0.6cm}|P{0.6cm}|P{0.6cm}|}
\hline
\textbf{Related Work} & \multicolumn{1}{c|}{\textbf{Year}} & \multicolumn{1}{c|}{\textbf{Covered Topics}}                                                                                  & \multicolumn{1}{c|}{\textbf{\begin{tabular}[c]{@{}c@{}}Data \\collection  \end{tabular}}} & \multicolumn{1}{c|}{\textbf{\begin{tabular}[c]{@{}c@{}} MOS \\ calc. \end{tabular}}} & \multicolumn{1}{c|}{\textbf{\begin{tabular}[c]{@{}c@{}} Human\\  factors\end{tabular}}} & \multicolumn{1}{c|}{\textbf{\begin{tabular}[c]{@{}c@{}} Utilize\\  ML\end{tabular}}} & \multicolumn{1}{c|}{\textbf{\begin{tabular}[c]{@{}c@{}}  End-to-end \\ \end{tabular}}} & \multicolumn{1}{c|}{\textbf{\begin{tabular}[c]{@{}c@{}}  ITU \\ P.1203  \end{tabular}}} & \multicolumn{1}{c|}{\textbf{\begin{tabular}[c]{@{}c@{}}  Helpful\\  for \\ RA\end{tabular}}}                                                                                                \\ \hline
Baraković \textit{et al.}\cite{f4-2}                   & 2013                               & Modeling, monitoring, measuring QoE                                                                         & \xmark                                                                                          & \xmark                                                                                          & \xmark                                                                                                   & \xmark                                                                              & \xmark                                                                                                                     & \xmark                                                                                 & \xmark                                                                                                     \\ \hline
Sultan \textit{et al.} \cite{f4-3}                     & 2023                               & Evaluation of multimedia services, based on QoE, communication networks with high bandwidth and low latency & \xmark                                                                                          & \xmark                                                                                          & \cmark                                                                                                   & \cmark                                                                              & \xmark                                                                                                                     & \cmark                                                                                 & \cmark                                                                                                     \\ \hline
Barman \textit{et al.} \cite{f4-4}                      & 2019                               & A review of QoE assessment models for adaptive video streaming                                              & \xmark                                                                                          & \cmark (Video params)                                                                       & \cmark                                                                                                   & \cmark                                                                              & \xmark                                                                                                                     & \xmark                                                                                 & \xmark                                                                                                     \\ \hline
Liotou \textit{et al.} \cite{f4-5}                      & 2023                               & The impact of using an intermediate server for caching on user QoE                                          & \xmark                                                                                          & \cmark (Video params)                                                                       & \cmark                                                                                                   & \xmark                                                                              & \xmark                                                                                                                     & \cmark                                                                                 & \cmark                                                                                                     \\ \hline
Barakabitze \textit{et al.} \cite{f4-6}                 & 2019                               & QoE management solutions for multimedia services in future networks                                         & \xmark                                                                                          & \xmark                                                                                          & \cmark                                                                                                   & \cmark                                                                              & \xmark                                                                                                                     & \xmark                                                                                 & \cmark                                                                                                     \\ \hline
Kougioumtzidis \textit{et al.} \cite{j1}              & 2022                               & Quality assessment in multimedia QoE and ML-based prediction                                  & \xmark                                                                                          & \cmark (Video params)                                                                       & \cmark                                                                                                   & \cmark                                                                              & \xmark                                                                                                                     & \xmark                                                                                 & \cmark                                                                                                     \\ \hline
Omar \textit{et al.} \cite{f4-7}                        & 2023                               & Using ML to predict QoE in multimedia networks                                                & 
\begin{tabular}[]{@{}c@{}}\cmark \\(Subj.)  \end{tabular}

                                                                             & \cmark (Net. params)                                                                     & \cmark                                                                                                   & \cmark                                                                              & \xmark                                                                                                                     & \xmark                                                                                 & \cmark

                                               \\ \hline

 \multicolumn{2}{|c|}{This research}                                 & Calculating QoE in Multimedia Services based on ML                                            & \cmark                                                                                          & \cmark (Obj.)                                                                              & \cmark (Net. params)                                                                              & \cmark                                                                              & \cmark                                                                                                                     & \cmark                                                                                 & \cmark

                                                                             \\ \hline
\end{tabular}
\label{tbl:related-works}

\end{table*}

These frameworks, available in both \textit{closed-source} and \textit{open-source} formats, are pivotal in evaluating and improving user satisfaction within communication networks. Closed-source frameworks, typically developed by private organizations, provide comprehensive solutions with proprietary features and dedicated support. In contrast, open-source frameworks, developed through community collaboration, offer transparency, flexibility, and customization options.
Reference \cite{sr-10} introduces an open-source Android application designed to assess YouTube user QoE by measuring network performance parameters and converting them into QoE scores. This application validates existing theoretical models through a pilot study that incorporates user feedback, leading to the development of a new empirical model based on the collected data. Another tool, \texttt{VLQoE}, presented in \cite{sr-14}, evaluates video QoE on smartphones and excels at predicting QoE by accurately simulating video stalls, making it valuable for QoE optimization studies.
Reference \cite{sr-16} explores an open-source framework that replicates end-user perception of mobile broadband services by simulating key quality indicators. Additionally, Reference \cite{sr-15} proposes an open-source model for assessing QoE in IoT multimedia services, introducing the pure boost score as a new QoE metric. Reference \cite{sr-17} presents proprietary machine learning (ML)-based QoE prediction models that utilize data gathered from a field trial in operational cellular networks. Similarly, Reference \cite{sr-20} describes a proprietary proactive LTE network management framework based on user QoE, employing various ML algorithms to accurately predict QoE parameters.

Mobile network operators are expected to manage the growing demand while maintaining a high video QoE. Achieving this objective requires operators to have a comprehensive understanding of the video QoE of users to support network planning, provisioning, and traffic management. However, several challenges arise when designing a system to measure video QoE:
\begin{itemize} \item The vast scale of video traffic data and the diversity of video streaming services, \item Multi-layer constraints stemming from the complex architecture of cellular networks, \item The difficulty in extracting QoE metrics from network traffic \cite{p1-2}, \item The challenge of achieving high confidence levels in QoE measurement due to factors such as the variety of terminal devices, diverse services, variations in media content, fluctuations in playback and network conditions, and significant spatial and temporal differences in device performance \cite{j1-7}. \end{itemize}
The success of a service depends significantly on user acceptance. Effective QoE management ensures end-user satisfaction by addressing their needs and expectations. Consequently, satisfied users are more inclined to adopt new and more complex services, thereby driving technological growth and advancement \cite{p1-3}.

\section{Contribution}

This paper presents an end-to-end framework for mobile network operators (MNOs) to predict and optimize the QoE for video streaming services based solely on network conditions, without requiring insight into the video content itself. The approach leverages ML to predict the mean opinion score (MOS) by training a random forest model on network KPIs such as delay, jitter, packet loss, throughput, and bitrate, in addition to per-segment MOS measurements computed using the ITU-T P.1203 standard.
A dataset was created by developing a video streaming data collection system using \texttt{Selenium}. This system extracts segment files from video streaming sessions, calculates per-segment MOS scores using the ITU P.1203 reference software, and stores them alongside network measurements such as delay and packet loss. Videos were streamed under various network conditions emulated on remote servers to capture a wide range of quality levels. In total, more than 20,000 labeled video segments were collected. The simulation results demonstrate that the random forest model can predict MOS from network metrics with a $R^2$ score of 0.958, which confirms the effectiveness of this approach for the optimization of multimedia QoE.
The key contributions of this research are as follows.
\begin{itemize} \item \textbf{End-to-End Framework:} This research introduces a framework that enables MNOs to optimize video streaming QoE using only network information. \item \textbf{Video Streaming Data Collection System:} A custom data collection system utilizing Selenium was developed to gather data on video quality under various network conditions, which is used to train the ML model. \item \textbf{ML Approach:} The paper proposes an ML model (random forest) to predict user satisfaction based on network KPIs. \item \textbf{High Accuracy:} The proposed method demonstrates high accuracy, with an $R^2$ score of 0.958, validating its effectiveness for optimizing multimedia QoE. \end{itemize}

\subsection{Paper Structure}
The remainder of the paper is organized as follows: First, in \S \ref{sec:Background}, related work in the field of QoE assessment for multimedia services is reviewed. Next, the proposed framework for predicting QoE parameters is introduced in \S \ref{sec:ProposedFramework}. In \S \ref{sec:Dataset}, the dataset used in our experiments is detailed. An overview of the evaluation platforms is provided in \S \ref{sec:EvaluationPlatforms} and \S \ref{sec:Simulation} presents the implementation results. Finally, in \S \ref{sec:Conclusion}, the paper is concluded with key findings, and directions for future research are suggested.

\section{Related Works}
\label{sec:Background}



As summarized in \autoref{tbl:related-works}, various studies in the field of QoE assessment for multimedia services have addressed challenges and proposed solutions. Baraković \textit{et al.} \cite{f4-2} focused on modeling, monitoring, and measuring QoE, laying the groundwork for further research. However, their study did not provide specific tools for data collection or MOS calculation, nor did it explore human factors or incorporate ML techniques. Building on this foundation, Sultan \textit{et al.} \cite{f4-3} evaluated multimedia services based on QoE in high-bandwidth, low-latency communication networks. They considered human factors and applied ML algorithms but did not provide tools for data collection or MOS calculation. Their findings were useful for resource allocation, and they utilized the ITU P.1203 standard for their evaluations.
Barman \textit{et al.} \cite{f4-4} reviewed QoE assessment models for adaptive video streaming, offering tools for MOS calculation based on video parameters and accounting for human factors. While they employed ML techniques, they did not provide a comprehensive framework that covers the entire process from data collection to QoE calculation, as seen in Sultan \textit{et al.} \cite{f4-3}. Liotou \textit{et al.} \cite{f4-5} investigated the impact of using an intermediate server for caching on user QoE, providing tools for MOS calculation and considering human factors. However, they did not use ML algorithms or provide data collection tools which marked a step back in terms of predictive capability compared to the work by Barman \textit{et al.} \cite{f4-4}. Their study, nonetheless, was insightful for resource allocation decisions and employed the ITU P.1203 standard, aligning more closely with Sultan \textit{et al.} \cite{f4-3}.

Barakabitze \textit{et al.} \cite{f4-6} proposed QoE management solutions for multimedia services in future networks, considering human factors and utilizing ML techniques. However, their study did not provide specific tools for data collection or MOS calculation similar to the limitation seen in Barman \textit{et al.} \cite{f4-4}.
Kougioumtzidis \textit{et al.} \cite{j1} focused on quality assessment in multimedia QoE and ML-based prediction, offering tools for MOS calculation and examining human factors. While they applied ML algorithms, they did not present a comprehensive framework that spans from data collection to QoE calculation. Their results were beneficial for resource allocation. Omar \textit{et al.} \cite{f4-7} employed ML to predict QoE in multimedia networks, providing subjective data collection tools and considering network parameters. They also examined human factors and used ML techniques but did not offer a complete framework encompassing data collection to QoE calculation or utilize the ITU P.1203 standard. Their findings were useful for resource allocation as effectively as Liotou \textit{et al.} \cite{f4-5}.

In contrast to these works, this research \cite{f4-8} aims to provide a comprehensive framework for calculating QoE in multimedia services using ML techniques. We have designed a testbed for objective data collection and MOS calculation, focusing on network parameters as key factors. The research also examines the impact of human factors and employs ML algorithms to predict QoE. A key feature of this study is its complete framework, which covers the entire process from data collection to QoE calculation. Additionally, it uses the ITU P.1203 standard to ensure the reliability and comparability of the results. The outcomes of this research are expected to be valuable for efficient resource allocation in multimedia networks. By addressing the limitations of previous works and providing a holistic solution, this research makes a significant contribution to the field of QoE assessment and management in multimedia services.

\section{Proposed Framework}
\label{sec:ProposedFramework}

\begin{figure}
	\centering
	\includegraphics[trim={0cm 1cm 0cm 2.5cm },clip,width=0.9\linewidth]{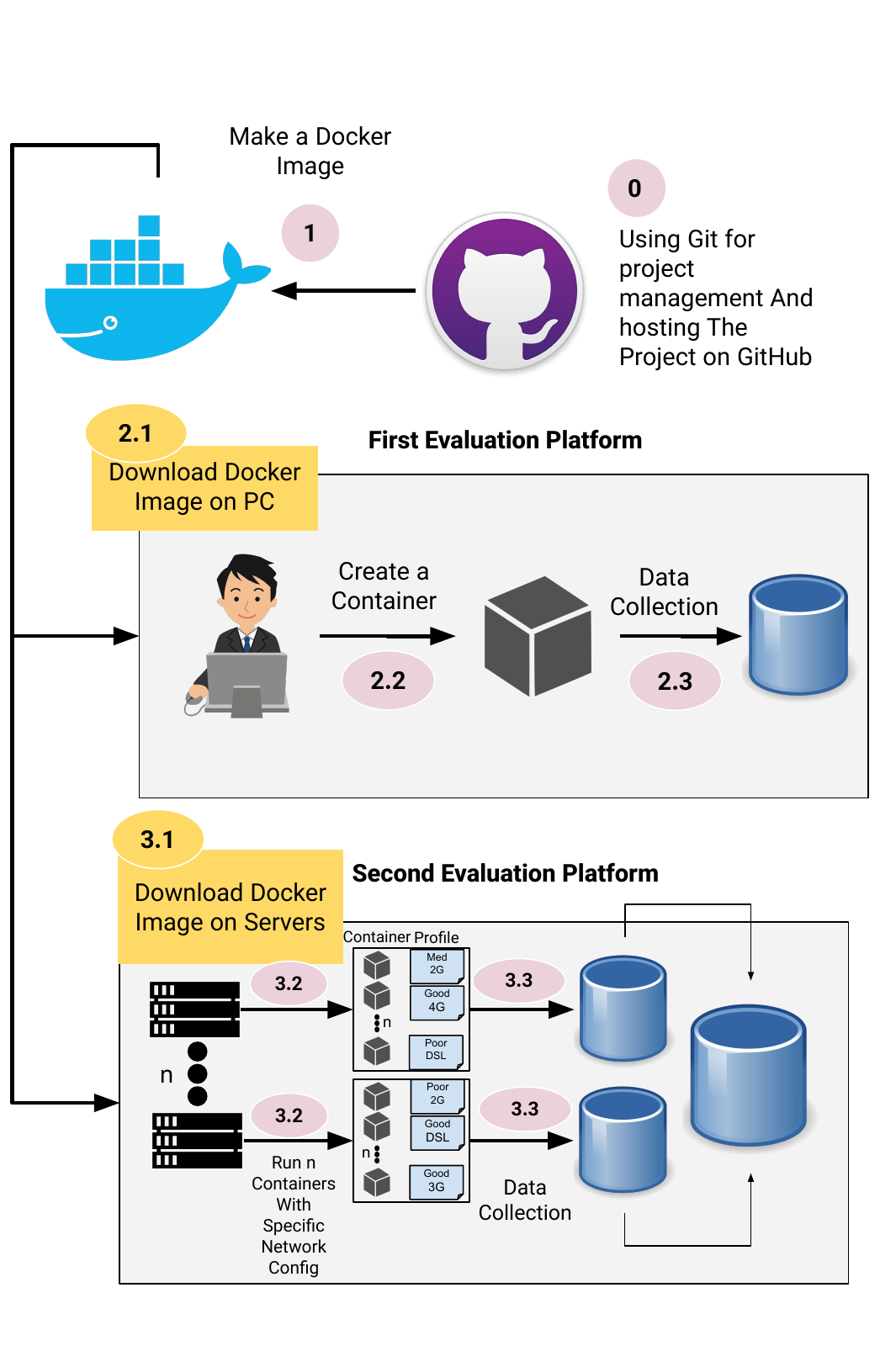}
	\caption[The implementation of the Data Collector Application]{Workflow of the data collection process. The initial evaluation platform involves downloading a Docker image onto a PC for data collection. The second evaluation platform expands this process to servers, enabling the simulation of diverse network configurations for comprehensive data analysis.}

	\label{fig:imp}
\end{figure}

This paper introduces an end-to-end framework for mobile network operators to predict and optimize QoE for video streaming services based solely on network conditions. 
As illustrated in Fig. \ref{fig:imp}, the execution and network simulation phase was conducted in two parts. The first part involved running the program on personal devices by various users with different Internet service providers. The second part consisted of simulations on servers located in different data centers, subjected to various network conditions. The implementation of this research is available as open-source software \cite{f4-8}, allowing other researchers and developers to use the project's source code, enhance it, or adapt it to their specific needs. This open-source approach fosters increased transparency and collaboration within the scientific community, facilitating further sharing of knowledge and experiences in the field.

We selected the ITU-T P.1203 model because of its suitability for short video segments in HTTP adaptive streaming. In the ITU P.1203 standard, video parameters are extracted and used as input to calculate the MOS for different aspects such as audio, video, and stalling. Our framework generates a standardized MOS that aligns with human perception of quality, as determined through subjective assessments. This method effectively labels video training data with quality scores, enabling the ML model to predict streaming satisfaction based solely on network metrics. The collected data is assigned continuous MOS values computed using the P.1203 standard.
The ITU-T software for calculating QoE accepts one or more audio/video files (segments) or input specifications in JSON format. Based on the input, the software computes per-second audio and video quality scores, as well as an overall integrated audiovisual quality score, in compliance with ITU P.1203 standards. Supported audio codecs include AAC-LC, HE-AAC, MP2, and AC-3, while the supported video format is H.264. The software autonomously determines the appropriate QoE calculation mode based on the input specifications:

\begin{enumerate}
    \item \textbf{Mode 0 (metadata only):} Uses bitrate, frame rate, and resolution.
    \item \textbf{Mode 1 (frame header data only):} Includes all items in Mode 0, plus frame types and frame sizes.
    \item \textbf{Mode 2 (two percent of bitstream data):} Includes all items in Mode 1, plus two percent of quantization parameter (QP) values of all frames.
    \item \textbf{Mode 3 (one hundred percent of bitstream data):} Includes all items in Mode 1, plus QP values of all frames.
\end{enumerate}

The QP is used to balance compression efficiency with video quality. A lower QP value results in less compression, higher video quality, and a larger file size. In contrast, a higher QP value results in more compression, lower video quality, and a smaller file size. According to this concept, higher modes offer greater prediction accuracy.

\begin{figure}
	\centering
	\includegraphics[trim={1cm 0cm 2.5cm 0cm },clip,width=0.9\linewidth]{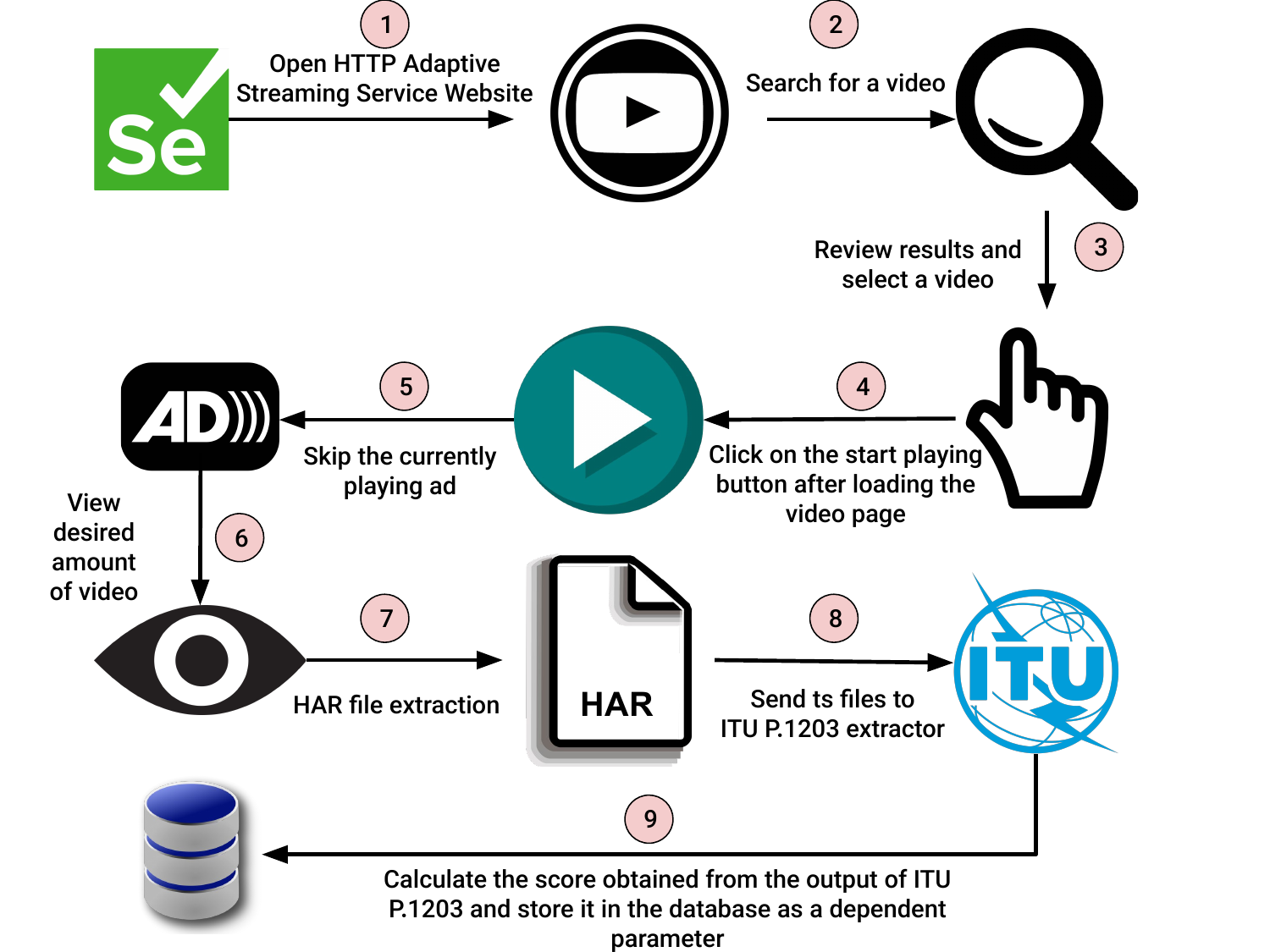}
	\caption{The data collection and MOS calculation process involves several steps. First, Selenium is utilized to access an HTTP Adaptive Streaming Service website and select a specific video. After skipping any advertisements, the desired video segment is played. Next, a HAR file is extracted to capture network traffic information. The extracted \texttt{.ts} video segments are then processed using the ITU P.1203 program to compute the QoE score. Finally, these scores plus network KPIs (\textit{e.g,} throughput, delay, jitter, packet loss, and bitrate) are stored in a database for further analysis.}
	\label{fig:final}
\end{figure}

\subsection{software application}
The software application of the proposed framework, which gathers video parameters and calculates quality can be divided into five modules: 

\begin{enumerate}
\item \textbf{Driver Initialization and Cleanup}:
The driver initialization module handles the setup of the WebDriver for Firefox. It installs the appropriate GeckoDriver based on the system architecture and initializes BrowserMob Proxy to intercept the network traffic. The cleanup module contains functions for cleaning up resources after the crawling process.
\item \textbf{Main Process Control}: This module contains the main logic for handling user input and initiating the crawling process. It includes the \textit{process\_input} function, which determines whether to process URLs from a command-line argument or a file.
\item  \textbf{Crawling and Data Collection}: This module handles the actual crawling process for a given URL.  It navigates to the URL, measures page load times and throughput, and interacts with the video player to detect advertisements and stalling events.
\item  \textbf{HAR Processing and Quality Assessment}: This module processes the HAR (HTTP Archive) file generated by BrowserMob Proxy. The \textit{process\_har} function extracts URLs of video segments from the HAR file and calculates QoS parameters such as startup time, buffering ratio, and average bitrate. It also calls a function to download and convert video segments for further analysis.
\item \textbf{Utility and Database Interaction}: It ensures that the collected metrics and QoE assessments are stored in a database for further analysis and reporting.
\end{enumerate}

To run the software, the user needs the following resources:
\begin{itemize}[label=$\ast$]
    \item Python version 3.7 or higher with pip3.
    \item To install dependencies, use the following code:
\begin{lstlisting}
pip3 install --user -r requirements.txt
\end{lstlisting}
    \item ffmpeg/ffprobe must be installed on the machine (only for mode 3).
\end{itemize}

\textbf{Input:}
As mentioned, there are two input formats: JSON and audiovisual segments. 
The files are provided to the software extractor using the following command, where the software extracts the necessary parameters from the files. The desired mode can also be specified at the end of the command.
\begin{lstlisting}
python3 -m itu\_p1203.extractor --use-average -m 3 /path/to/segment1.mp4 /path/to/segment2.mp4 > mode3.json
\end{lstlisting}

\textbf{Output:} Assuming a file is given as input to the software, the output is as follows (example):
\begin{lstlisting}
{
  "path/to/first/input/file": {
    "O21": [
      // per-second audio quality scores
    ],
    "O22": [
      // per-second video quality scores
    ],
    "O23": 5.0, // stalling quality
    "O34": [
      // per-second audiovisual quality scores
    ],
    "O35": 4.63, // audiovisual quality score
    "O46": 4.92, // overall quality score
    "mode": 0, // used mode, either 0, 1, or 3
    "streamId": 42 // currently unused
  },
  "path/to/second/input/file": {
    ...
  }
}
\end{lstlisting}
As can be seen, the output of this software is a JSON file. The output values printed by the software are as follows:

\textbf{"O21"}: per-second audio quality scores,

\textbf{"O22"}: per-second video quality scores,

\textbf{"O23"}: stalling quality score,

\textbf{"O34"}: per-second audiovisual quality scores,

\textbf{"O35"}: audiovisual quality score,

\textbf{"O46"}: overall quality score,

\textbf{"streamId"}: currently unused.

\subsection{Data collection apporach}

The purpose of this is to obtain labeled data and use it in an ML model to train this model. This program can be distributed and executed on user devices, and the ML model can also be trained in a distributed way:

\begin{enumerate}
	\item Execute the data collection program on personal devices under different real network conditions or simulate various network conditions on servers and collect data,
	\item Preprocess collected data then, select features like delay, jitter, packet loss, bit rate, throughput (independent variables) and MOS (dependent variable),
	\item Train models like linear regression, DNNs, random forest on 80\% of data and 20\% test data.
	
\end{enumerate}
As shown in Fig. \ref{fig:final}, a cycle of the first and second phases illustrates the data collection process's executive steps:

\begin{enumerate}[label=\Roman*.]
    \item \textbf{Accessing the Video streaming platform}:
    Utilize Selenium WebDriver to navigate to the designated video streaming platform's website.

    \item \textbf{Searching for the video}:
    Employ the search functionality provided by the platform to locate the specific video of interest.

    \item \textbf{Selecting the desired video}:
    Identify and select the video from the search results to initiate playback.

    \item \textbf{Initiating playback}:
    Locate the video start button and perform a click action to commence video playback.

    \item \textbf{Handling advertisements}:
    In the event that an advertisement plays, utilize Selenium to interact with the interface and skip the advertisement, ensuring uninterrupted access to the video content.

    \item \textbf{Watching the Video}:
    Observe the video for a duration specified in the startup configuration file, ensuring that the observation period aligns with the required assessment criteria.

    \item \textbf{Extracting the HAR file}:
    After the specified video playback duration, extract the HTTP Archive (HAR) file from the WebDriver session using Selenium's capabilities.

    \item \textbf{Processing the HAR file}:
    \begin{enumerate}[label*=\arabic*.]
        \item \textbf{Extraction of .ts Files}: Use regular expressions (Regex1) to identify and isolate the .ts (MPEG transport stream) files within the HAR file.
        \item \textbf{Downloading and Storing}: Download each identified .ts file and save them into a designated folder.
        \item \textbf{QoE Score Calculation}: Submit each .ts file individually to the ITU-T P.1203 program to compute the QoE score.
    \end{enumerate}

    \item \textbf{Data Storage and Labeling}:
    Append the computed QoE scores as labels to the corresponding video data and store the annotated data in a database for further analysis and retrieval.
\end{enumerate}

As part of the data collection process, we also measure KPIs such as throughput, delay, jitter, packet loss, and bitrate associated with the video streaming sessions.
\textit{Stalling} in video playback refers to interruptions that occur during video or film streaming, temporarily halting playback and disrupting the seamless display of content. These interruptions can happen unexpectedly and vary in length, negatively affecting the viewer experience and the MOS. The causes of stalling can be varied:

\begin{enumerate}
    \item \textbf{Internet connection weakness}: Stalling often results from weak or unstable internet connections. If the internet speed is insufficient for continuous video playback, stalling is likely to occur.

    \item \textbf{Network issues}: Problems within the communication network, whether in local networks or broader internet systems, can lead to stalling.

    \item \textbf{High network traffic}: High levels of network traffic, particularly during peak usage times or in areas with limited infrastructure, increase the likelihood of stalling.

    \item \textbf{Server issues}: Technical problems or insufficient capacity on video streaming servers can cause stalling.

    \item \textbf{User device technical problems}: Technical issues with the viewer's device, such as software or hardware malfunctions, can also contribute to stalling.
\end{enumerate}

To mitigate the stalling, improvements can be made such as enhancing internet connectivity, using services with stronger server capabilities, and even employing advanced technologies like content delivery network.
To calculate stalling using the ITU P.1203 standard, stalling events need to be identified in the form of an array [start time of stall, duration of stall]. Two methods can be employed for this purpose:

\begin{figure}
	\centering
	\includegraphics[trim={2.5cm 1cm 2.5cm 1cm},clip,width=0.95\linewidth]{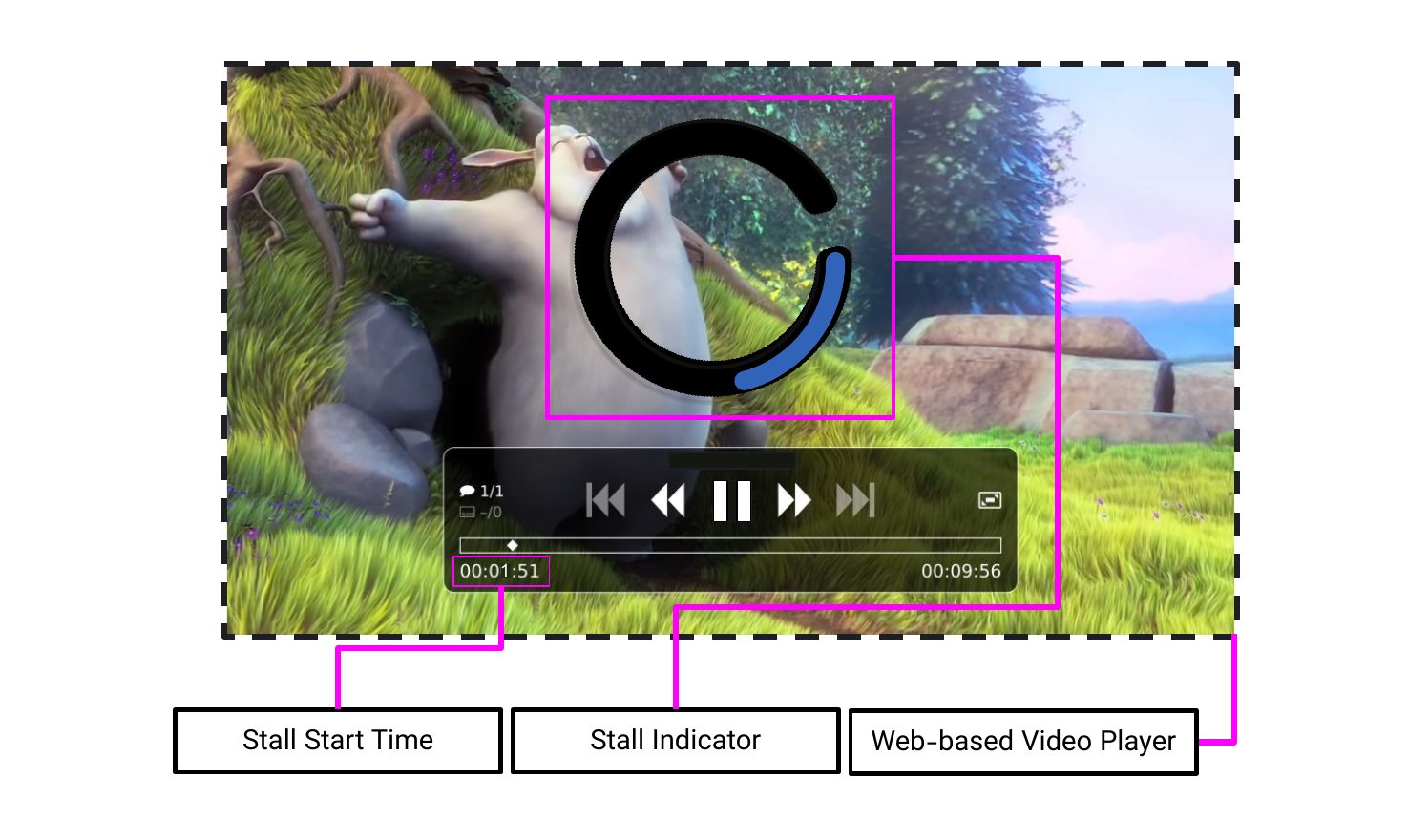}
	\caption[Stall example]{An example of a stall occurring within a web video player. The stall, indicated by the stall element, denotes a temporary interruption during the playback of content. Such stalls can vary in duration, ranging from brief pauses to more prolonged interruptions, significantly impacting the viewer's experience.}

\label{fig:stall}
\end{figure}

\textbf{1- Real-time calculation during playback:} 
In this method, a stall detector function within the application is activated during video playback. This function monitors the video player every second. As depicted in Fig. \ref{fig:stall}, a web video player is shown playing an animation and experiences a stall at 1:51, indicating that subsequent video segments have not been sent yet. The stall detector function records the start time of the stall as soon as it detects the stall indicator. It continues to monitor the player every second. If the stall indicator is still present, it increments the stall duration by one second. Otherwise, it populates the array \texttt{[start time of stall, duration of stall]} and stores it as the stalling parameter for future calculations using the ITU P.1203 standard.

\textbf{2- Calculation from HAR files:}
In this method, based on HAR files, the time of request dispatch for each video segment is specified, denoted by $T_{start}$. Three other times are defined for each specified request:
\texttt{$t_s$}: The time length for sending the desired video segment request,
\texttt{$t_w$}: The time duration to wait for an answer that remains,
\texttt{$t_r$}: The time length for the desired video segment to arrive at the player from the server.

According to the \autoref{form:segrec}, for the time a video segment n reaches the player, it is expressed as:
			\begin{equation}
				\label{form:segrec}
				\text{$T_{seg}$} = T_{start} +  t_s + t_w + t_r
			\end{equation}
Continuing, the length of time for each video segment is denoted by $d_n$ Now, if $T_{seg_{n+1}}$ exceeds the sum of the lengths of all segments up to segment \texttt{n}, it means that segment \texttt{n+1} has arrived after the completion of previous segments. Therefore, stalling has occurred.

According to this proposition, \autoref{form:sdetect} follows:
\begin{equation}
				\label{form:sdetect}
				\text{$T_{seg_{n+1}}$} > \sum_{i=1}^{n} \text{d}_i 
\end{equation}
Therefore, if this condition occurs, stalling has occurred, and the stall duration $S_d$ for segment \texttt{n+1}, denoted by equation \autoref{form:sresult}, is:

\ignore{
			\begin{equation}
				\label{form:sresult}
				S_{\text{d}_{n + 1} } =
				\begin{cases}
					T_{\text{seg}_{n+1}} - \sum_{i=1}^{n} ( \text{d}_i +  \text{$S_{d}$}_i )& \text{If  } T_{\text{seg}_{n+1}} > \sum_{i=1}^{n} (\text{d}_i + \text{$S_{d}$}_i)\\
					0 & \text{otherwise}
				\end{cases}
			\end{equation}
}

\begin{equation}
    \label{form:sresult}
    S_{\text{d}_{n+1}} =
    \begin{cases}
        T_{\text{seg}_{n+1}} - \sum_{i=1}^{n} (\text{d}_i + S_{\text{d}_i}) & \text{if } T_{\text{seg}_{n+1}} > \\
        \quad \sum_{i=1}^{n} (\text{d}_i + S_{\text{d}_i}) \\
        0 & \text{O.W}
    \end{cases}
\end{equation}

Therefore, the total stall duration is given by equation \autoref{form:ssum}.
\begin{equation}
				\label{form:ssum}
				\text{$S_{total}$} > \sum_{i=1}^{n} \text{S}_{d_i}
			\end{equation}

The algorithm introduced in this research for stall calculation relies on the previous method, as implementing it requires knowing the duration of each video segment. This approach is chosen because calculating segment durations adds computational cost and complexity.

\section{Dataset}
\label{sec:Dataset}
Table \ref{tbl:db} presents a detailed snapshot of 15 rows from the database, summarizing key parameters collected during video playback tests. The table condenses information such as video URLs, test times, and test numbers, while providing crucial metrics like throughput (bps), estimated satisfaction (MOS), video resolution (width and height), video duration, frame rate, startup time, buffering time, total size with buffer, bitrate (kbps), delay, jitter, and packet loss percentage. These metrics are vital for analyzing the quality of video streaming experiences and evaluating performance and user satisfaction as per the ITU P.1203 standard.
\begin{table*}[]
\caption{Detailed snapshot of 15 rows from the database.}

\begin{adjustbox}{width=\textwidth}
\begin{tabular}{|
>{\columncolor[HTML]{F8A102}}l |
>{\columncolor[HTML]{DAE8FC}}l |
>{\columncolor[HTML]{DAE8FC}}l |
>{\columncolor[HTML]{DAE8FC}}l |
>{\columncolor[HTML]{DAE8FC}}l |
>{\columncolor[HTML]{DAE8FC}}l |l|l|l|l|
>{\columncolor[HTML]{F8FF00}}l |l|l|l|}
\hline
{\color[HTML]{000000} mos} & loss & jitter & delay & bitrate & throughput & rebuffering & buffering & framerate & duration & stalling         & vheight & vwidth & startup \\ \hline
{\color[HTML]{000000} 242} & 1000 & 43     & 66    & 310     & 28680      & 4580        & 1780      & 3404      & 50000    & 6 - 10           & 360     & 640    & 920     \\ \hline
{\color[HTML]{000000} 191} & 0    & 83     & 242   & 660     & 60034      & 0           & 4730      & 2065      & 41000    & 7 - 10           & 264     & 264    & 2040    \\ \hline
{\color[HTML]{000000} 293} & 0    & 54     & 78    & 820     & 61616      & 0           & 2650      & 4940      & 40000    & 7 - 10           & 480     & 854    & 750     \\ \hline
{\color[HTML]{000000} 206} & 3000 & 86     & 116   & 290     & 11315      & 3340        & 1570      & 1700      & 42000    & 3 - 20 | 7 - 10  & 264     & 264    & 2320    \\ \hline
{\color[HTML]{000000} 216} & 0    & 90     & 135   & 400     & 32850      & 0           & 3140      & 6628      & 49000    & 8 - 10           & 146     & 264    & 570     \\ \hline
{\color[HTML]{000000} 199} & 2000 & 81     & 48    & 150     & 41440      & 0           & 1910      & 2337      & 45000    & 8 - 10           & 148     & 264    & 2070    \\ \hline
{\color[HTML]{000000} 459} & 1000 & 133    & 415   & 540     & 47065      & 0           & 1200      & 3753      & 41000    & 8 - 10           & 720     & 1280   & 270     \\ \hline
{\color[HTML]{000000} 219} & 0    & 81     & 57    & 550     & 37532      & 3230        & 1390      & 3736      & 49000    & 8 - 10           & 360     & 640    & 680     \\ \hline
{\color[HTML]{000000} 254} & 0    & 59     & 81    & 710     & 48481      & 0           & 1280      & 4132      & 40000    & 6 - 15 | 8 - 10  & 426     & 426    & 3570    \\ \hline
{\color[HTML]{000000} 215} & 0    & 132    & 177   & 430     & 17397      & 0           & 2170      & 3535      & 45000    & 15 - 20 | 8 - 10 & 240     & 426    & 1610    \\ \hline
{\color[HTML]{000000} 169} & 0    & 104    & 47    & 250     & 126440     & 0           & 1060      & 3921      & 42000    & 9 - 10           & 240     & 426    & 1380    \\ \hline
{\color[HTML]{000000} 258} & 0    & 44     & 61    & 270     & 41767      & 0           & 3600      & 7871      & 46000    & 9 - 10           & 360     & 640    & 2180    \\ \hline
{\color[HTML]{000000} 306} & 0    & 121    & 3     & 410     & 40390      & 0           & 1830      & 3732      & 52000    & 9 - 10           & 480     & 854    & 770     \\ \hline
{\color[HTML]{000000} 328} & 0    & 67     & 51    & 810     & 34687      & 0           & 1010      & 3778      & 41000    & 1 - 20 | 9 - 10  & 480     & 854    & 670     \\ \hline
{\color[HTML]{000000} 306} & 0    & 89     & 2     & 410     & 14726      & 3130        & 1830      & 4343      & 50000    & 0 - 0            & 240     & 426    & 740     \\ \hline

\end{tabular}
    \end{adjustbox}

\label{tbl:db}
\end{table*}
\ignore{
\autoref{tbl:db} illustrates a sample of 15 rows from the database, with many actual parameters (such as video URL, test time, test number, etc.) omitted for brevity. Some of the columns in the table are explained as follows:

\begin{itemize}
\item \textbf{throughput:} Throughput measured in bits per second.
\item Estimated MOS according to ITU P.1203, multiplied by 100.
\item \textbf{vwidth and vheight:} Indicating the resolution of the video.
\item \textbf{duration:} The duration for which the video was watched (modifiable according to settings, but set to 40 seconds in this study).
\item \textbf{framerate:} The frame rate, multiplied by 100.
\item \textbf{startup:} The start-up time of playback, measured in milliseconds (ms).
\item \textbf{buffering:} Buffering time, measured in milliseconds (ms).
\item \textbf{bitrate:}, measured in kilobits per second (kbps).
\item \textbf{delay:} Delay obtained from ping, measured in milliseconds (ms).
\item \textbf{jitter:} Jitter obtained from ping, measured in milliseconds (ms).
\item \textbf{loss:} Packet loss percentage obtained from ping, multiplied by 100.
\end{itemize}
}
As illustrated in Table \ref{tbl:db}, the columns are color-coded to differentiate between various types of parameters:

\begin{enumerate}
    \item \textbf{Uncolored Columns}: These columns represent parameters not involved in training the ML model. these parameters were collected to maintain the independence of each section of the program from others. These parameters may be required in other multimedia quality research (e.g., image resolution).
    \item \textbf{Yellow Columns}: The yellow-highlighted column indicates stalling, with cells showing 0-0 denoting no stalling. For example, in row five, two stalling events are recorded: the first stall occurs at the 3-second mark of the video and lasts for 20 seconds, and the second stall starts at the 7-second mark and lasts for 10 seconds.
    \item \textbf{Blue Columns}: These columns represent parameters used as independent variables in training the ML model. These parameters are network-related, reflecting the study’s approach to using these metrics for estimating user satisfaction and resource allocation. Acquiring these parameters from users is simpler for operators compared to obtaining video-related parameters, allowing operators to estimate user satisfaction effectively with fewer network parameters compared to numerous and complex video-related parameters.
    \item \textbf{Orange Columns}: The orange column represents the MOS parameter, which is the target variable. This column is involved as the dependent parameter in the ML model training. For instance, in row one, the score of 2.42 indicates that, according to the ITU P.1203 standard, this video has been assigned an MOS of 2.42. It should be noted that this MOS score may increase with better network conditions and decrease with worse network conditions.
\end{enumerate}
Some numbers in the database have been scaled by a factor of 100 to prevent storing floating-point numbers and to facilitate storage as integers, thereby improving the quality of the database. Furthermore, the data set used in this research has been uploaded to Kaggle \cite{kaggle-data}, making it accessible for further analysis and reproduction of the study results by other researchers.



\section{Evaluation platforms}
\label{sec:EvaluationPlatforms}
To thoroughly assess the performance and adaptability of the proposed framework, evaluations were conducted across different platforms, each designed to simulate various real-world and controlled environments. The evaluation process was divided into two main platforms, which are detailed in the following subsections:

\subsection{First Evaluation Platform}

\begin{figure}
	\centering
	\includegraphics[trim={0cm 0cm 0cm 0cm},clip,width=0.99\linewidth]{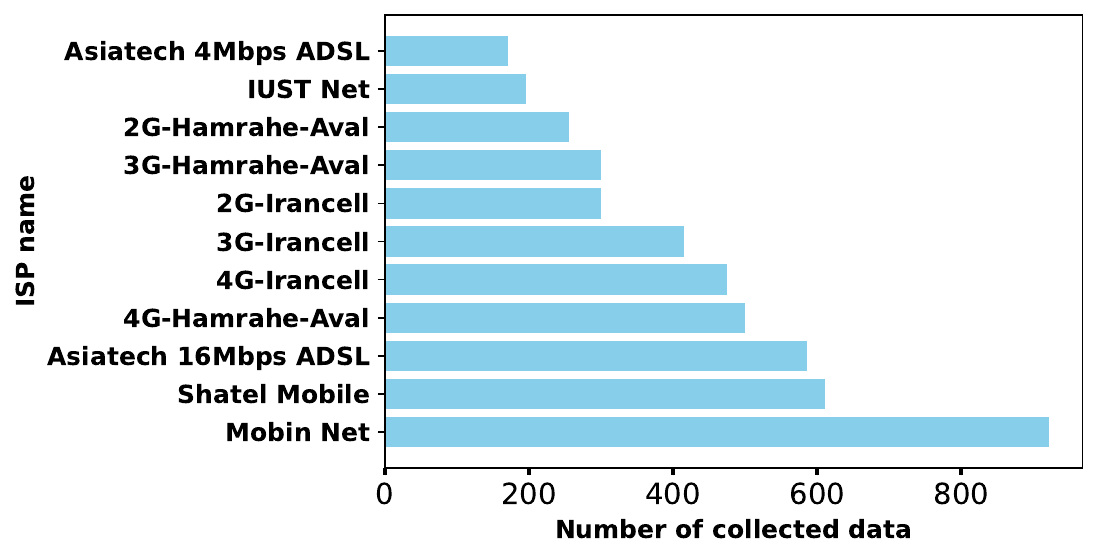}
	\caption[Results of the First Evaluation Platform]{Results of the First Evaluation Platform, showing the number of collected data points across various ISPs. The chart highlights the diversity of network conditions evaluated, with Mobin Net and Shatel Mobile providing the highest data collection volumes.}

\label{fig:personalimp}
\end{figure}

As shown in Fig. \ref{fig:imp}, the program was executed on individuals' personal computers. The evaluation was actually performed on the personal computers of 8 people to examine the effects and performance of the program in different environments with various devices, operating systems, and Internet service providers. This assessment included performance, stability, and compatibility with different devices and services (Fig. \ref{fig:personalimp}). These personal computers were running Windows, Linux, and macOS operating systems. All of these computers had appropriate and sufficient hardware to execute the program.

\subsection{Second Evaluation Platform}

\begin{figure*}
	\centering
	\includegraphics[trim={1cm 0.75cm 0cm 0cm},clip,width=0.8\linewidth]{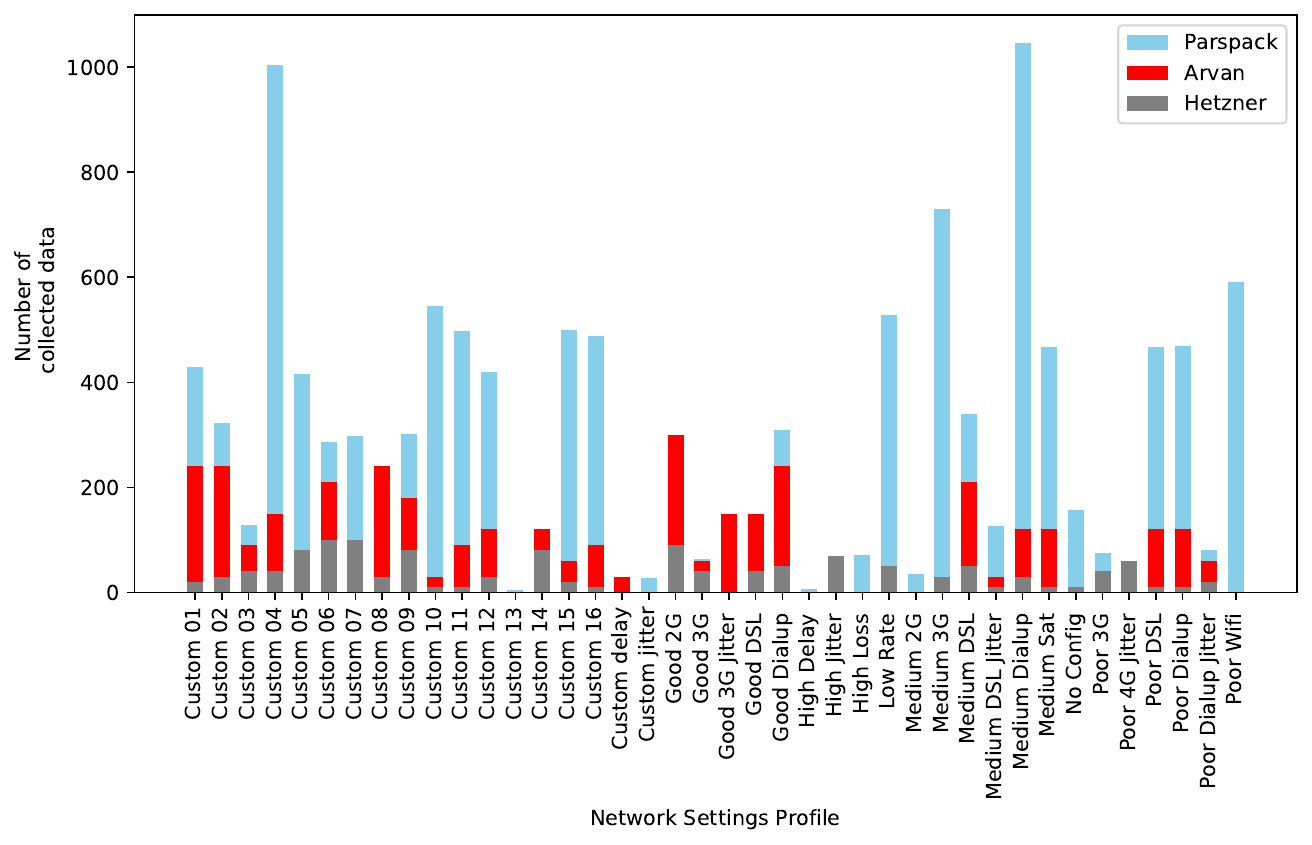}
	\caption[Results of the Second Evaluation Platform]{Results of the Second Evaluation Platform, illustrating the number of collected data points across various network settings profiles. The data was gathered from three different cloud service providers: ParsPack, Arvan, and Hetzner. Each provider's data is represented by different colors: blue for ParsPack, red for Arvan, and gray for Hetzner. The chart demonstrates a diverse range of network conditions, including custom configurations and standard profiles such as Good 3G, Poor 4G, and High Jitter. The variation in collected data highlights the robustness of the evaluation process and the comprehensive coverage of different network scenarios, essential for accurate QoE assessment.}
	\label{fig:serverimp}
\end{figure*}

As shown in Fig. \ref{fig:imp}, in the second execution platform, containers were running on several servers. The evaluation was performed on Arvan Cloud, Parspack, and Hetzner Cloud servers. Each server was equipped with 8 GB of RAM, 4 CPU cores with amd64 architecture. Four Docker containers were running on each server, with each container configured with different network parameters. This setup allows for the simulation of various real network conditions and users, enabling the assessment of MOS scores for the observed videos under these conditions. Fig. \ref{fig:serverimp} illustrates the data collection from different servers with various network configuration profiles.

The \texttt{config.txt} file defines various network profiles to simulate different conditions such as bandwidth, delay, jitter, and packet loss. Each profile represents a specific network environment, such as 4G, 3G, Wi-Fi, or custom scenarios with varying levels of performance. Standard profiles represent common network types (e.g., Good 4G, Poor 3G). Each profile specifies values for incoming and outgoing network parameters, while custom profiles allow for highly specific network conditions tailored to particular needs, such as high jitter or low bandwidth. These profiles facilitate the simulation of unusual or extreme network scenarios.

The core idea is to use the \texttt{config.txt} file to define and adjust network conditions for simulation. This setup enables controlled testing environments where various network profiles can be systematically tested. By gathering performance data under these simulated conditions, you can analyze how different network environments impact MOS.

An example profiles:
\begin{lstlisting}
// Good 4G Network Profile:
-incoming
delay 20ms
delay-distro 5ms
loss 0%
rate 10Mbps
-outgoing
delay 20ms
delay-distro 5ms
loss 0%
rate 5Mbps
\end{lstlisting}

\subsection{Implementation and System Requirements}
The implementation of this research uses Python 3.11 and an \texttt{sqlite3} database for simplicity and speed. The maximum RAM requirement is 700 MB, and 1 CPU core is sufficient to run the program. The disk space usage by the program code is less than 100 MB; however, an additional storage buffer needs to be considered for the video file currently being processed by the program. Therefore, allocating 500 MB of disk space for this purpose is suitable.
All systems running on AMD and ARM architectures are suitable to execute the program. Building the code for other architectures is also possible. The tested operating systems to run the program are: Windows (7 and 10), Linux, and macOS (both Intel and Apple Silicon architectures). The software execution test was also performed on the Raspberry Pi 5 single-board computer running the Raspberry OS.
Therefore, in general, the program is capable of running and collecting data on all operating systems that support Docker (e.g, Android). \texttt{Git} was also used to manage the project, and the project was uploaded as open source on 
GitHub \cite{im-1}.

\subsection{Learning model}
\autoref{ml-specs} presents a comparison of three ML models employed for predicting the relationship between independent and dependent variables: Linear Regression, Random Forest, and Deep Neural Network (DNN). Linear Regression serves as a baseline model, capturing the linear relationship between variables. The Random Forest model, based on decision trees, combines multiple trees to enhance prediction accuracy. It is characterized by parameters such as the number of trees (\texttt{n\_estimators}), maximum depth of each tree (\texttt{max\_depth}), and maximum features used in each node (\texttt{max\_features}). On the other hand, the DNN model utilizes a multilayer perceptron architecture with one input layer, four hidden layers, and one output layer. The hidden layers employ the ReLU activation function, while the output layer uses a linear activation function. The DNN model is trained using the Adam optimizer and mean absolute error (MAE) as the loss function, with a specified number of training epochs and batch size. These models offer diverse approaches to capture and predict the relationships between variables, each with its own set of specifications and parameters.

The required data on the first and second platforms was collected using the data collection system with Selenium, resulting in approximately 20,000 rows stored in the database. Next, we proceed to the training phase of the ML model. The goal of this phase is to train a ML model using network-related parameters, enabling the determination of the MOS using only these network parameters. Initially, the collected data is processed by removing problematic entries. Then, processes such as feature engineering and normalization are applied to the data. Consequently, features such as delay, jitter, packet loss, bitrate, and throughput are selected as independent variables, while MOS serves as the dependent variable for model training.

\begin{table}[]

\caption{Specifications and parameters of the ML models used in this study. The models include Linear Regression, Random Forest, and DNN, each with detailed configurations to predict the relationship between independent and dependent variables}

\begin{tabular}{|p{2cm}|p{6cm}|}
\hline
\multicolumn{1}{|c|}{\textbf{Model}} & \multicolumn{1}{c|}{\textbf{Specifications and Parameters}}                                                                                                                                                                                                                                                                                                                                          \\ \hline
Linear Regression                    & A baseline model for predicting the linear relationship between independent and dependent variables                                                                                                                                                                                                                                                                                                  \\ \hline
Random Forest                        & A decision tree-based model                                                                                                                                                                                                                                                                                                                                                                          \\ \hline
Random Forest (continued)            & \begin{tabular}[c]{@{}l@{}}Combination of multiple decision trees \\ to improve prediction accuracy: \\ -Number of trees (n\_estimators): 600\\ -Maximum depth of each tree (max\_depth): 48\\ -Maximum features in each node (max\_features): 0.58\end{tabular}                                                                                                                                  \\ \hline
Deep Neural Network (DNN)            & \begin{tabular}[c]{@{}l@{}}A neural network-based model using a multilayer \\ perceptron architecture\\ -Network Architecture:\\ -One input layer with 128 neurons\\ -Four hidden layers with 256 neurons in each layer\\ -One output layer with one neuron\\ Activation Functions:\\ -Hidden layers: ReLU\\ -Output layer: Linear\\ Number of training epochs: 2000\\ Batch size: 32\end{tabular} \\ \hline
Deep Neural Network (continued)      & \begin{tabular}[c]{@{}l@{}}Optimizer: Adam\\ Loss function: Mean Absolute Error (MAE)\end{tabular}                                                                                                                                                                                                                                                                                                   \\ \hline
\end{tabular}
\label{ml-specs}
\end{table}

\section{Implementation results}
\label{sec:Simulation}
In this section, we describe the methodology implemented to predict the MOS for QoE using various network parameters. The implementation follows a systematic approach comprising data loading, cleaning, feature engineering, data visualization, and model training and evaluation.
Initially, the dataset is loaded and cleaned to ensure the quality of the data. Rows with anomalous values for \texttt{delay\_qos}, \texttt{avg\_bitrate}, and \texttt{jitter} are removed or adjusted to prevent these anomalies from skewing the analysis. Specifically, rows with \texttt{delay\_qos} values of -1000 or 0, and \texttt{avg\_bitrate} values of 0 are dropped. Similarly, \texttt{jitter} values of 0 or -1000 are adjusted to 1 to avoid logarithmic transformation issues. After cleaning, 20,411 data points are retained for analysis.

\subsection{Feature Engineering}
To enhance the predictive power of the model, several new features are engineered:
\begin{itemize}
    \item \textbf{Logarithmic Transformations:} Logarithmic transformations of \texttt{delay\_qos} and \texttt{avg\_bitrate} are created to normalize their distributions.
    \item \textbf{Interaction Terms:} Interaction terms such as \texttt{throughput * jitter} and \texttt{delay * jitter} are introduced to capture the combined effects of these parameters.
    \item \textbf{Polynomial Features:} A squared term for \texttt{packet\_loss} is included to account for its nonlinear effects.
    \item \textbf{Rate of Loss:} The ratio of \texttt{avg\_bitrate} to \texttt{packet\_loss} is calculated to create a \texttt{loss\_rate} feature.
\end{itemize}
These new features aim to reduce the mean squared error (MSE) and enhance the model's predictive accuracy.

\subsection{Data Visualization}
To understand the relationships between the features and the target variable MOS, scatter plots are created for each new feature against MOS. This visualization step helps in verifying the linearity and significance of the new features.

\subsection{Model Training and Evaluation}
Two ML models, Linear Regression and Random Forest Regressor, are used to predict MOS. The data is split into training and testing sets to evaluate the models' performance.

\begin{enumerate}
    \item \textbf{Baseline Model:} Initially, the models are trained using the original features (\texttt{delay\_qos}, \texttt{avg\_bitrate}, \texttt{jitter}, \texttt{throughput}, \texttt{packet\_loss}). The performance metrics such as MSE, Root Mean Squared Error (RMSE), R-squared (R\textsuperscript{2}), and Mean Absolute Error (MAE) are recorded. The baseline model yielded the following results:

    \begin{itemize}
        \item \textbf{Old MSE:} 2.8306e-05
        \item \textbf{New RMSE:} 0.0053
        \item \textbf{R-squared:} 0.6252
        \item \textbf{MAE:} 0.0039
    \end{itemize}

    \item \textbf{Enhanced Model:} Subsequently, the models are trained using the new engineered features. A Random Forest Regressor with optimized hyperparameters (e.g., number of estimators, max depth, max features) is utilized to improve prediction accuracy. The performance metrics are recalculated and compared with the baseline model. The enhanced model demonstrated improved performance with the following results:

    \begin{itemize}
        \item \textbf{New MSE:} 3.1047e-06
        \item \textbf{New RMSE:} 0.0018
        \item \textbf{R-squared:} 0.9589
        \item \textbf{MAE:} 0.0012
    \end{itemize}
\end{enumerate}

The enhanced model demonstrates improved performance, as evidenced by lower MSE and RMSE values, and higher R\textsuperscript{2} scores, indicating a better fit to the data.


\begin{figure}
    \centering
   \includegraphics[trim={1.2cm 1.2cm 1.2cm 1.5cm},clip,width=0.99\linewidth]{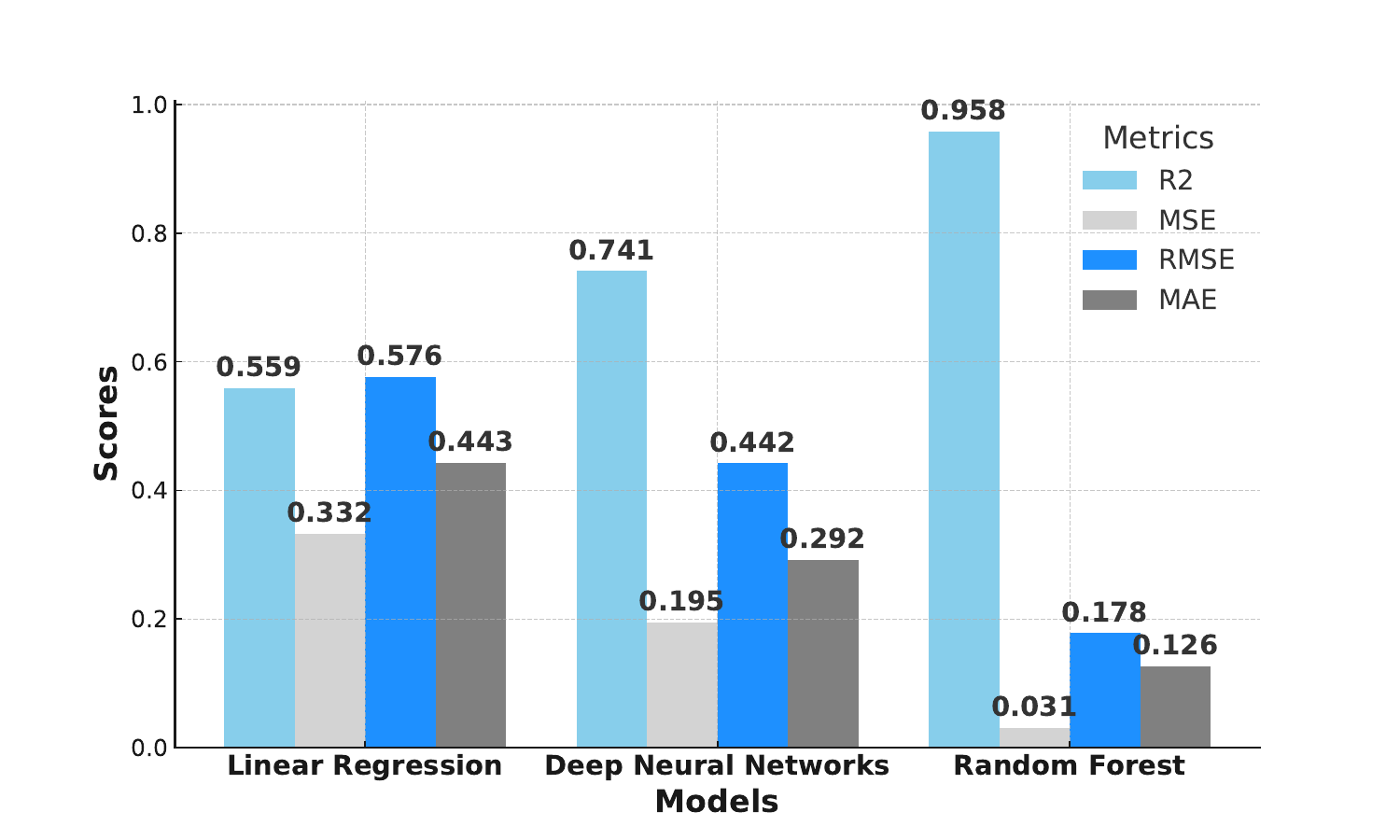}
  \caption{This figure provides a detailed comparison of the predictive accuracy of Linear Regression, Deep Neural Networks, and Random Forest models across various quality metrics which include, R2, MSE, RMSE, and MAE.}
  

    \label{fig:results-chart}
\end{figure}

As shown in Fig. \ref{fig:results-chart}, three models of linear regression, deep neural networks, and random forest were trained with 80\% of the obtained data, then tested with the remaining 20\% data. As a result, according to the obtained results, it can be determined how accurately the model can estimate the MOS (the dependent variable of the model) given the mentioned network-related inputs.
According to the obtained values, the Random Forest model yielded the best result with the highest $R^2$ value of 0.958 and the lowest error, which can be considered as the key research finding. Therefore, given the tools introduced for collecting video streaming data and calculating MOS in phases 1 and 2, as well as network simulation and program execution in phase 3 of the implementation, the required data for phase 4, which involved ML model training, was obtained. 
The Random Forest model was then trained with 80\% of the data and tested with 20\% of the data to validate its performance. According to the obtained numerical results, a reliable model was achieved for predicting MOS by using only five main network parameters: delay, jitter, packet loss, bit rate, and throughput. Thus, this research provides a framework for calculating the QoE of video streaming multimedia services based on ML.

\section{Conclusion and future works}
\label{sec:Conclusion}
This research set out to develop a robust framework for predicting and optimizing the QoE in multimedia services, with a specific focus on video streaming applications. The primary idea was to address the growing need for real-time, accurate QoE assessments that can inform network resource allocation and improve user satisfaction. By leveraging machine learning techniques, particularly the Random Forest algorithm, we aimed to predict the MOS using readily available network parameters such as delay, jitter, packet loss, bitrate, and throughput. These parameters were chosen because they are indicative of the user’s experience and can be monitored without needing to access the actual content being streamed. To ensure that the predictions were closely aligned with human perceptions of quality, we based our QoE assessments on the ITU-T P.1203 standard, which provides a rigorous framework for evaluating audiovisual quality. This approach not only enhances the precision of QoE predictions but also contributes to a standardized methodology that can be adopted by other researchers and industry practitioners.

The implementation of this framework involved an extensive data collection process, wherein over 20,000 video segments were streamed under varied network conditions to capture a broad spectrum of quality experiences. These segments were then labeled with MOS values calculated using the ITU P.1203 standard, providing a rich dataset for training and testing the machine learning models. Our results demonstrated that the Random Forest model outperformed other models, achieving an impressive R\textsuperscript{2} value of 0.958, indicating a high level of accuracy in predicting QoE based solely on network parameters. This finding is particularly significant because it validates the hypothesis that network-centric metrics can reliably predict user satisfaction, thus enabling network operators to dynamically optimize resource allocation without needing to process or analyze the content itself. Such an approach reduces computational overhead and enhances the scalability of QoE optimization strategies in diverse network environments.

There are several avenues for expanding and refining this framework in the future. One potential direction is the integration of more sophisticated machine learning models, such as deep learning, which could further enhance the accuracy and robustness of QoE predictions by capturing more complex patterns in the data. Additionally, the framework could be adapted for use in other types of multimedia services, such as social networking platforms or messaging apps, where QoE plays a critical role in user engagement. Another promising area of exploration is the application of distributed learning techniques, such as federated learning, which would allow the model to be trained across multiple devices without centralized data collection, thus preserving user privacy while continuously improving the model's performance. By pursuing these and other future directions, the framework could evolve into a versatile tool for ensuring high levels of user satisfaction across a wide range of digital services, thereby supporting the ongoing growth and evolution of the digital economy.

\appendices

\bibliographystyle{IEEEtran}
\bibliography{References.bib}

\end{document}